# NMR Evidence of Charge Fluctuations in Multiferroic CuBr$_2$

Rui-Qi Wang(王瑞琦)[1,2], Jia-Cheng Zheng(郑家成)[2], Tao Chen(陈涛)[2], Peng-Shuai Wang(王朋帅)[2], Jin-Shan Zhang(张金珊)[3], Yi Cui(崔祎)[2], Chao Wang(王超)[4], Yuan Li(李源)[4], Sheng Xu(徐胜)[1], Feng Yuan(袁峰)[1] and Wei-qiang Yu(于伟强)[2†]

[1]*College of Physics, Qingdao University, Qingdao 266071, China*
[2]*Department of Physics and Beijing Key Laboratory of Opto-electronic Functional Materials & Micro-nano Devices, Renmin University of China, Beijing 100872, China*
[3]*Mathematics and Physics Department, North China Electric Power University, Beijing 102206, China*
[4]*International Center for Quantum Materials, School of Physics, Peking University, Beijing 100871, China*

We report combined magnetic susceptibility, dielectric constant, nuclear quadruple resonance (NQR) and zero-field nuclear magnetic resonance (NMR) measurements on single crystals of multiferroics CuBr$_2$. High quality of the sample is demonstrated by the sharp magnetic and magnetic-driven ferroelectric transition at $T_N = T_C \approx$ 74 K. The zero-field $^{79}$Br and $^{81}$Br NMR are resolved below $T_N$. The spin-lattice relaxation rates reveal charge fluctuations when cooled below 60 K. Evidences of an increase of NMR linewidth, a reduction of dielectric constant, and an increase of magnetic susceptibility are also seen at low temperatures. These data suggest an emergent instability which competes with the spiral magnetic ordering and the ferroelectricity. Candidate mechanisms are discussed based on the quasi-one-dimensional (1D) nature of the magnetic system.

## 1. Introduction

Multiferroic materials with both magnetic ordering and ferroelectricity are promising candidates for applications in mutual control of magnetism and ferroelectricity [1,2].This research is heated by the discovery of materials with magnetoelectric couplings [3–7]. Unfortunately, applications in these materials are still limited either because of weak magnetoelectric coupling, or their low transition temperatures. In case of type-II multiferroics, where strong magnetoelectric coupling is manifest by the magnetic driven ferroelectricity [8], their transition temperatures $T_N$ (= $T_C$) usually set in at fairly low temperatures. Understanding of magnetoelectric coupling and searching for tuning method to enhance the $T_N$ in type-II material are still on the way. In this aspect, broader investigation of material properties are strongly urged in this field [9].

In addition to most oxide type-II multiferroics, CuCl$_2$ was only discovered in

This Work is supported by the Ministry of Science and Technology of China (Grant Nos. 2016YFA0300504), the National Science Foundation of China (Grant Nos.11374364), and the Fundamental Research Funds for the Central Universities and the Research Funds of Renmin University of China (Grant No. 14XNLF08).
†Corresponding author. E-mail:wqyu_phy@ruc.edu.cn

2009 as a non-oxide material, which has large electrical polarization below the antiferromagnetic (AFM) transition temperature $T_N \approx 24$ K [10,11], Though its transition temperature is fairly low, it has promoted exploration of new multiferroics on its sister compounds. A recent discovery is the multiferroicity in $CuBr_2$, which has the same distorted $CdI_2$ structure as $CuCl_2$ (see Fig. 1(a)), with a high transition temperature $T_N$ ($T_C$) of 73.5 K and a low loss charge character [12]. Magnetic frustration with quasi-1D $J_1 - J_2$ interactions and the inverse Dzyaloshinskii-Moriya (DM) interaction seem to play an important role for the emergent spiral magnetic ordering and the magnetic-driven ferroelectricity [13–15].

NMR studies have been reported on several oxide multiferroics, but mostly limited to the magnetic properties [16,17]. $CuBr_2$ provides a superior system to study the magnetoelectric and/or magnetoelectric couplings, because $^{79}Br$ and $^{81}Br$ are $S = 3/2$ nuclei with large natural abundances and quadrupole moments, which is suitable for both NQR and NMR measurements. Indeed, several NMR and NQR studies have been reported on $CuBr_2$ [18–21], although mostly focusing on properties above $T_N$.

Here we report the combined zero-field NQR (above $T_N$) and NMR (below $T_N$) studies to reveal the interplay of magnetism and charge properties. We first resolved the $^{79}Br$ and $^{81}Br$ NMR spectra. A large anomalous enhancement of the spin-lattice relaxation rates $1/^{79}T_1$ and $1/^{81}T_1$ is seen when cooled far below $T_N$. Surprisingly, our detailed analysis on both nuclear sites resolves that the enhanced $1/T_1$ is primarily caused by charge fluctuations, rather than magnetic fluctuations. Meanwhile, we found that the enhanced charge fluctuations are accompanied with the abnormal increases of the NMR linewidth, the reduction of the dielectric constant, and the increase of the magnetic susceptibility. These phenomena reveal an anti-correlation between charge fluctuations and the magnetic ordering, and suggest a competing ground state for this quasi-1D magnetic system.

The paper is organized as following. In Section 2, we show the crystal structure and measurement techniques. In Section 3, the high quality of the single crystals is demonstrated by the sharp magnetic and ferroelectric transition at about 74 K, with evidences of magnetoelectric coupling. The high-temperature NQR data are briefly shown in Section 4. In Section 5, we present the zero-field NMR spectra, with the assignment of the $^{79}Br$ and the $^{81}Br$ resonance peaks. Details data and analysis on the spin-lattice relaxation rates are given in Section 6. Discussions with charge fluctuations are presented in Section 7, with a short summary given in Section 8.

## 2. Materials and techniques

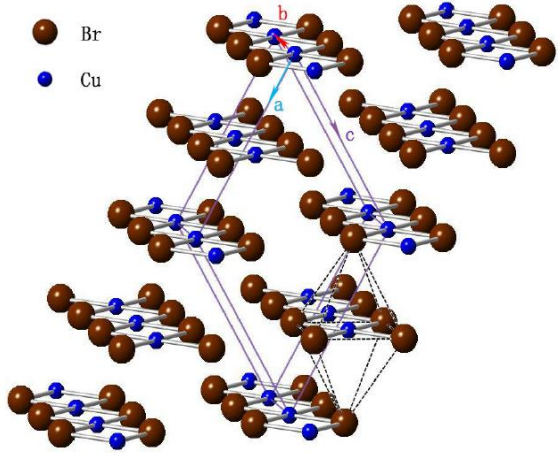

**Fig. 1.** The crystal structure of CuBr$_2$. $a$, $b$, and $c$ are axis of the monoclinic structure of the crystal. CuBr$_2$ ribbons are directed along the crystalline $b$-direction to form the quasi-1D structure. Br$^-$ ions in three adjacent ribbons forms elongated octahedron, proximately along the $a$ direction.

As shown in Fig. 1, Cu$^{2+}$ spins are strongly coupled along the $b$-axis to form a 1D CuBr$_2$ ribbon. The couplings along the $a$ and $c$ axis are much weaker, because of much large lattice parameters of $a$ and $c$. Along the ribbon, ferromagnetic exchange coupling $J_1$ is induced among nearest Cu$^{2+}$ neighbors from the Kramers theorem, because the adjacent Cu-Br-Cu bond has a nearly 90° angle. On the other hand, antiferromagnetic exchange coupling $J_2$ is formed between next nearest neighbors. Such competing interactions lead to the spiral magnetic ordering, which also causes magnetic-driven ferroelectricity due to DM interactions [12].

The single crystals of anhydrous CuBr$_2$ were grown by slow evaporation of aqueous solutions [22]. The crystals are plate like with a dimension of 10*10*1 mm$^3$, with cleavage surfaces along the $ab$ plane of the crystal. The magnetic susceptibility is measured by a VSM in a 14 T PPMS (Quantum Design). The ferroelectric measurement is performed by a capacitive method. A pair of conducting plates are attached to two cleavage surfaces of a single crystal, and the capacitance between two plates is measured by a LCR meter (Agilent 4263B).

For the NQR and NMR studies, the sample is cut into a size of 3 × 5 × 1 mm$^3$. All NQR and NMR measurements are performed under zero field, taking advantage of the EFG (electric-field-gradient) above $T_N$ and the static internal hyperfine field in the magnetically ordered state. The spectra are accumulated by the standard spin-echo sequence $\pi/2 - \tau - \pi$, with $\pi$ pulses ~ 1$\mu s$ and $\tau \approx 4\mu s$. The spin-lattice relaxation times, $^{79}T_1$ and $^{81}T_1$, are measured by the inversion-recovery method, where the magnetization data are fit to the standard recovery functions for $S = 3/2$ spins. We found only a single component of $T_1$ with no stretching behaviors in the magnetization recovery, in the presented $T_1$ data, which rules out extrinsic contributions.

## 3. Magnetic susceptibility and dielectric constant measurements

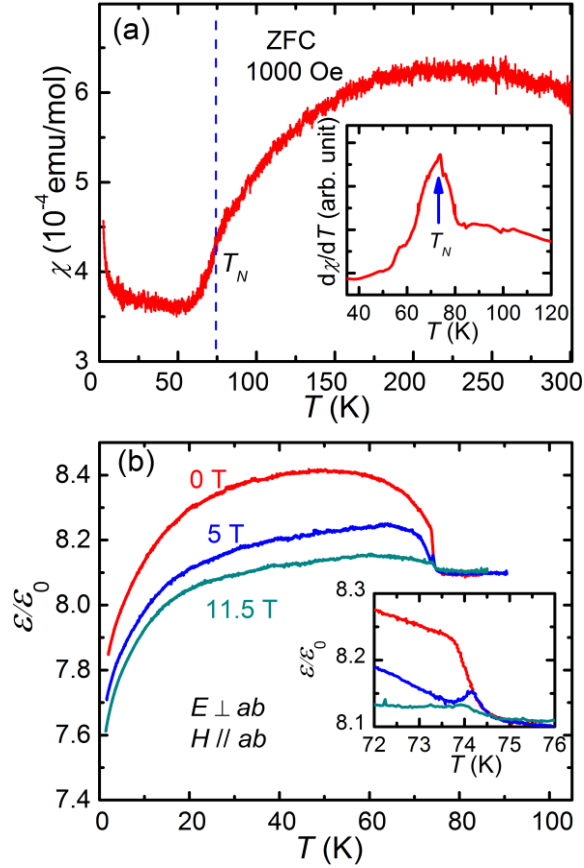

**Fig. 2.** (color online) (a) The magnetic susceptibility $\chi(T)$ of CuBr$_2$ as a function of temperature, with a 1000 Oe magnetic field applied in the $ab$-plane, under zero-field-cooling (ZFC) condition. Inset: the temperature derivative, $d\chi/dT$. The blue arrow points at a sharp peak characterizing the magnetic transition at $T_N$. (b) The $ac$ dielectric constant $\varepsilon$ of the sample measured under different magnetic fields, with capacitance measurement frequency 20 kHz. The magnetic field are applied along the $ab$-plane of the crystal. Inset: The enlarged view of the data close to transition temperature $T_N$.

Both the magnetic and dielectric measurements are performed on CuBr$_2$ single crystals. The low-field magnetic susceptibility $\chi$ is measured with field along the $ab$ plane, with data shown in Fig. 2(a). The magnetic transition into the spiral magnetic ordering is seen at $T_N$, by a small kinked feature in $\chi$. The $T_N$ is precisely determined to be 74 K by a sharp peaked feature in the temperature derivative of $\chi(T)$, as shown in the inset of Fig. 2(a). This $T_N$ is slightly higher than earlier reports [12].

The dielectric constant along the $c$-direction is obtained by calculating $\varepsilon = cd/S$, where $c$ is the capacitance between two copper plates attached the sample, $d$ is the thickness of the sample, and $S$ is the surface area of the copper plates. As shown in Fig. 2(b), $\varepsilon$ demonstrates a sharp increase just at $T_N$ ($\sim$ 74 K), evidencing the magnetic-driven ferroelectric transition [12]. Compared with powder samples, our ferroelectric transition exhibits a very rapid increase just below $T_N$. In fact, from the inset of Fig. 2(b), the transition width is only about 0.5 K, indicating very high quality of our single crystals.

The magnetoelectric coupling is further demonstrated by the change of $\varepsilon$ under magnetic field applied in the $ab$-plane, as shown in the inset of Fig. 2(b). With field increased from 0 to 5 T, a large decrease of $\varepsilon$ is clearly seen below $T_N$. With further increase of field up to 11.5 T, the change of $\varepsilon$ becomes weak. This is consistent with a spin-op transition at about 5 T [12]. With this field orientations, the spin-op transition rotates the magnetic moment with a $c$-axis component, and thus reduces the charge polarization along the same direction. This observation should be a strong support for the mechanism of magnetic-driven ferroelectricity [6,7].

Below 50 K, a small uprise in $\chi$ and a reduction in $\varepsilon$ are clearly seen with decreasing temperature. Since the onset temperature is very high, these anomalous behaviors should be attributed to an intrinsic cause of our high quality samples. This will be further discussed in Section 7.

## 4. NQR measurements above $T_N$

In $CuBr_2$, four types of isotopes, $^{63}Cu$, $^{65}Cu$, $^{79}Br$, and $^{81}Br$ are available for NQR/NMR studies. Their nature abundance, gyromagnetic ratio $\gamma$, quadrupole moments Q, and $\nu_Q$ in $CuBr_2$ at the ambient conditions are shown in Table.1.

**Table 1.** NQR characteristics for $CuBr_2$. The $\nu_Q$ listed here are measured at 293 K [19].

|  | spin | abundance(%) | $\gamma_n$(MHz/T) | Q($10^{-28}m^2$) | $\nu_Q$(MHz) |
|---|---|---|---|---|---|
| $^{63}Cu$ | 3/2 | 69.09 | 11.285 | -0.21 | 30.37 |
| $^{65}Cu$ | 3/2 | 30.91 | 12.099 | -0.195 | 28.18 |
| $^{79}Br$ | 3/2 | 50.54 | 10.667 | 0.33 | 83.98 |
| $^{81}Br$ | 3/2 | 49.46 | 11.499 | 0.28 | 70.10 |

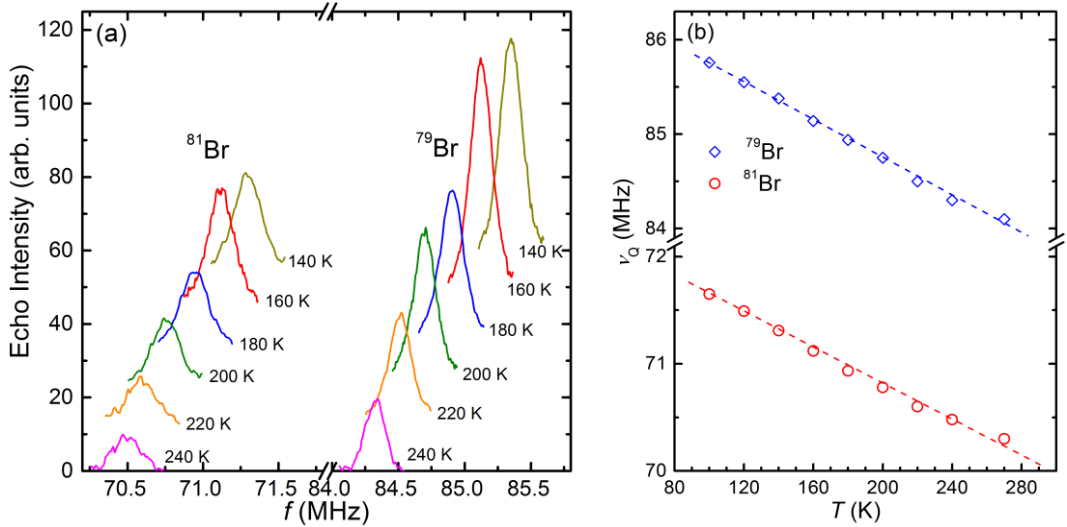

**Fig. 3.** (color online) (a) The NQR spectra of $^{79}Br$ and $^{81}Br$, at typical temperatures above $T_N$.

Data are offset for clarity. (b) The $^{79}\nu_Q$ and $^{81}\nu_Q$ as functions of temperature determined by the NQR measurement above $T_N$.

We measured the NQR spectra of $^{79}$Br and $^{81}$Br above $T_N$, whose resonance frequency are determined by local electric-field-gradient (EFG). In Fig. 3(a), the NQR spectra of $^{79}$Br and $^{81}$Br are shown at typical temperatures. The spectrum of each type of nucleus has a single line at all temperatures above $T_N$. Their resonance frequencies are shown as functions of temperature in Fig. 3(b), consistent with earlier reports [18,19,21]. Both $^{79}\nu_Q$ and $^{81}\nu_Q$ follow a linear increase with decreasing temperature, indicating a progressive lattice contraction.

## 5. Zero-field NMR measurement below $T_N$

When the temperature drops below $T_N$, zero-field NMR can be performed, since static hyperfine field is produced by the ordered moments of $Cu^{2+}$. Both the local hyperfine field and the local EFG act on the nucleus [23] as expressed below:

$$\mathcal{H} = -\gamma\hbar \mathbf{I} \cdot \mathbf{H_{hf}} + \frac{eQV_{zz}}{4I(2I-1)}[(3I_z^2 - I^2) + \eta(I_x^2 - I_y^2)]. \tag{1}$$

Here $\mathbf{H_{hf}}$ is the local hyperfine field produced by electrons, $V_{zz}$ is the EFG tensor, and $\eta = (V_{xx} - V_{yy})/V_{zz}$ is the asymmetric factor of local EFG.

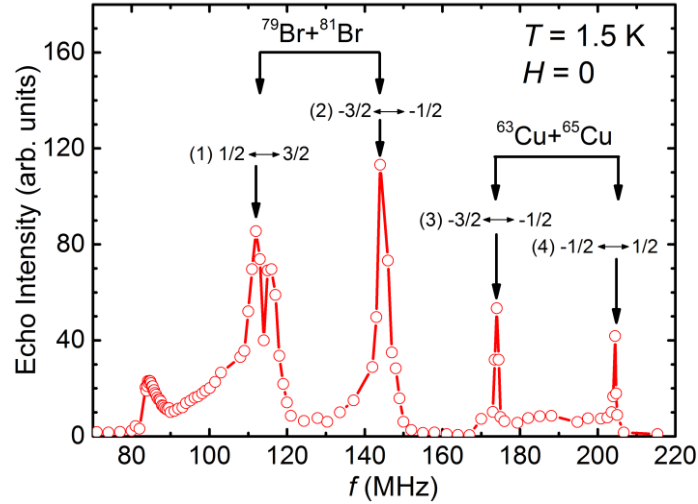

**Fig. 4.** The zero-field NMR spectrum of $CuBr_2$ measured at 1.5 K and in the frequency window from 70 to 220 MHz.

In all, twelve zero-field NMR lines are expected in the ordered state, considering that $^{63}$Cu, $^{65}$Cu, $^{79}$Br and $^{81}$Br are spin-3/2 isotopes. Fig. 4 shows a scan of zero-field NMR spectra in the measured frequency range from 70 to 220 MHz, at a fixed temperature 1.56 K. Here, four sets of NMR lines are clearly seen, labeled as (1) to (4). In order to assign all these spectra, we consider the following facts. i) $^{63}$Cu and $^{65}$Cu have close values of gyromagnetic ratios and the quadrupole moments, which would lead to their NMR spectra with close frequencies. ii) $^{79}$Br and $^{81}$Br also have close values of gyromagnetic ratios and quadrupole moments. Therefore, their NMR spectra should also have close frequencies. iii) $^{63}$Cu and $^{65}$Cu has much larger gyromagnetic ratios than that of $^{79}$Br and $^{81}$Br. $^{63}$Cu and $^{65}$Cu should have much larger internal fields in the ordered state than that of $^{79}$Br and $^{81}$Br, because the ordered moments are on the Cu ions. iv) $^{63}$Cu and $^{65}$Cu has much smaller quadrupolar frequencies than that of $^{79}$Br and $^{81}$Br, as already shown in the high-temperature data in Table 1.

With points i) and ii), six groups of lines are expected for zero-field NMR in the ordered state, including -1/2 ↔ 1/2, -3/2 ↔ -1/2, and 1/2 ↔ 3/2 transitions for Cu and Br respectively. Point iii) leads to a much higher zero-field NMR resonance frequency for $^{63}$Cu and $^{65}$Cu, and point iv) causes a narrower NMR linewidth for $^{63}$Cu and $^{65}$Cu, than that of $^{79}$Br and $^{81}$Br. Therefore, we attribute line (1) and (2) to the Br signals, and (3) and (4) to Cu signals. Another two lines are missing, which likely fall out of our resolution window and/or frequency window respectively, with one Br spectra below 60 MHz and one Cu spectra above 235 MHz.

Line (4) is reasonably described as the center transition (-1/2 ↔ 1/2) of $^{63}$Cu and $^{65}$Cu. Here we can't separate $^{63}$Cu and $^{65}$Cu lines. $^{63}$Cu has a larger quadrupole moment $Q$ but a smaller $\gamma$ than $^{65}$Cu, and the combination of both effects may lead to the mixing of two lines. Line (3) is about 30 MHz lower and has a similar line shape to line (4), which is consistent with $^{63}$Cu / $^{65}$Cu satellite lines (-3/2 ↔ -1/2), with $^{63}\nu_Q$ and $^{65}\nu_Q$ close to 30 MHz as shown in Table. 1. From this, the (1/2 ↔ 3/2) is estimated to be at 237 MHz, out of our measurement range.

We assigned low-frequency (80 ~ 160 MHz) peaks as $^{79}$Br and $^{81}$Br lines with different transitions, as denoted in Fig. 4. Lines (1) and (2) are broad, simply because of the incommensurate magnetic structure and the large EFG of the $^{79}$Br and $^{81}$Br sites, where the second-order EFG contribution broadens the spectra largely. The (-1/2 ↔ 1/2) is estimated to be below 60 MHz. We did not find an apparent peak at such low frequencies, possibly because of weak signal to noise ratio.

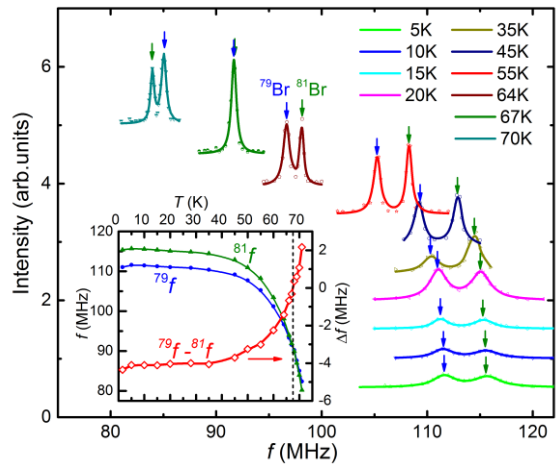

**Fig. 5.** Zero-field $^{79}$Br and $^{81}$Br NMR spectra at typical temperatures below $T_N$. Data are offset for clarity. Inset: The resonance frequencies $^{79}f$ and $^{81}f$ deduced by the peaked positions, shown as functions of temperature, with scales on the left axis. Their differences $^{79}f - ^{81}f$ are shown with scales on the right axis.

The relative assignment of the $^{79}$Br and $^{81}$Br peaks in the line (1) is also supported by their nearly equal peak height, given that the natural abundance of both isotopes are about 50%. We further assign the two peaks in line (1) as respective $^{79}$Br and $^{81}$Br lines as noted in Fig. 5. The temperature dependence of $^{79}$Br and $^{81}$Br resonance frequencies follows a combined effect of hyperfine field and EFG contributions as shown below. The spectra of both isotopes with escalating temperatures are shown in Fig. 5. The two peaks are well resolved at low temperature, and then mixed into a single peak at 67 K. Above 67 K, the two-peak feature reemerges. This is understood by the fact that $^{81}$Br has a $\gamma$ 8% higher than that of $^{79}$Br, but a 15.2% lower $Q$. With temperature increasing toward $T_N$, the hyperfine field decreases because of the reduction of the ordered moment of $Cu^{2+}$, and the EFG contribution becomes relatively larger. This is exactly demonstrated in the spectra: at low temperatures, a larger resonance frequency is seen on the $^{81}$Br nuclei because of dominant hyperfine contributions, but a higher resonance frequency is seen on the $^{79}$Br nuclei close to $T_N$ when the EFG has a larger contribution. Close to $T_N$, all NMR signals are wiped out due to slow magnetic fluctuations.

In principle, a double peak feature should also been seen in the line (2). However, the line (2) has a broad linewidth of 10 MHz, and it is not surprising if the $^{79}$Br and $^{81}$Br lines are mixed in this set of spectra for real materials, even though a double line feature is seen in the line (1). Our relative assignment of two peaks of line (1) to $^{79}$Br and $^{81}$Br is further supported by $1/T_1$ data later, because this assignment gives dominant spin fluctuations close to $T_N$ as theoretically expected. Any other assignment will not lead to this effect. Nevertheless, the main conclusion of this article is based on the relative assignment of $^{79}$Br and $^{81}$Br of the line (1), and we will not further discuss line (2), (3) and (4).

In the inset of Fig. 5, the respective resonance frequencies $^{79}f$ and $^{81}f$, and the difference between them, are plotted as functions of temperatures. When temperature goes up close to $T_N$, a rapid drop in $^{79}f$ and $^{81}f$ are seen, which follows the decrease of the ordered magnetic moments. $^{79}f - ^{81}f$ increases rapidly and crosses zero, due to increased EFG contributions as described above. At 71 K, the two resonance frequencies are about 80.2 MHz and 82.37 MHz, slightly higher than the $^{79}\nu_Q$ and $^{81}\nu_Q$ at 80 K, but much higher than $^{63}\nu_Q$ and $^{65}\nu_Q$ (see Table. 1), again supporting our assignment of the $^{79}$Br and the $^{81}$Br NMR lines.

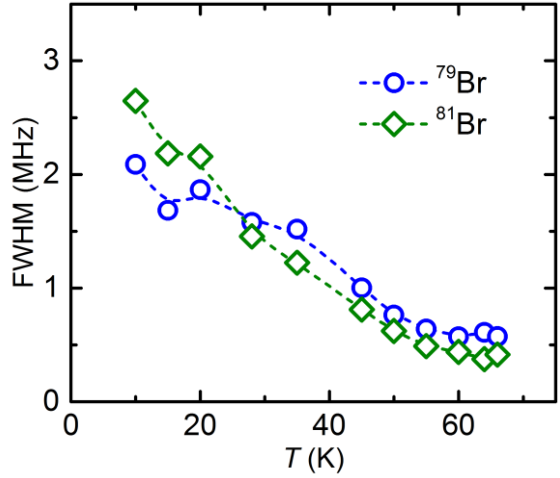

**Fig. 6.** The FWHM of the $^{79}$Br and the $^{81}$Br NMR spectra, as functions of temperature.

Another prominent observation is that both NMR peaks broaden significantly upon cooling. In Fig. 6, the Full-width-at-half-maximum (FWHM) of both peaks are plotted as functions of temperature. A large increase of the FWHM emerges when cooled below ∼ 50 K, which does not saturate even at 10 K. By contrast, the peak frequencies $^{79}f$ and $^{81}f$, presented in Fig. 5, increase sharply when cooled below 70 K, and level off with temperature below 50 K. The contrasting behaviors between the center frequency and the NMR linewidth indicates the formation of local magnetic inhomogeneity, which will be discussed in Section 7.

## 6. Spin-Lattice Relaxation Rates

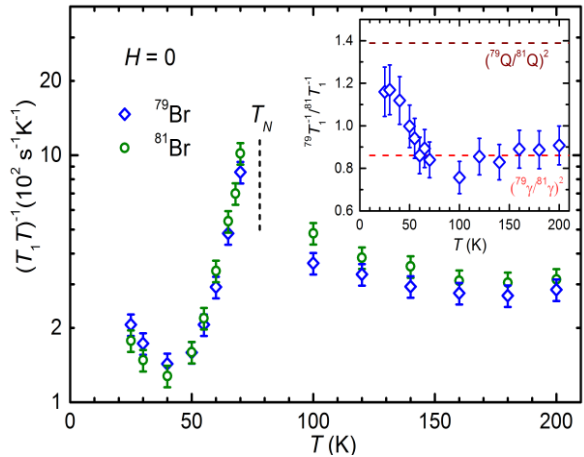

**Fig. 7.** The spin-lattice relaxation rates divided by temperature, $1/T_1T$, measured by ZF NMR (below $T_N$) and by NQR (above $T_N$) respectively. Inset: The ratio of the $1/T_1$ of two types of isotopes.

With the assignment of two NMR lines to $^{79}$Br and $^{81}$Br, we are now ready to study the low-energy fluctuations by the spin-lattice relaxation rates. As shown in Fig. 7, the spin-lattice relaxation rates divided by temperature, $1/T_1T$, are shown for both $^{79}$Br and $^{81}$Br under zero-field conditions. As we can see from Fig. 7, a peaked feature in $1/T_1T$ is seen at $T_N$ for isotopes, consistent with a critical slowing down behavior at the magnetic transition.

When the sample is cooled below $\sim 40$ K, an upturn arises in $1/T_1T$, which suggests the onset of low-energy spin fluctuations, even though the temperature is far below $T_N$. The ratio between $1/^{79}T_1$ and $1/^{81}T_1$, as shown in the inset of Fig. 7, also varies with temperature. $^{79}T_1^{-1}/^{81}T_1^{-1}$ is a constant from 200 K down to 60 K, and then increases upon further cooling. The change of the ratio is a result of varying contributions from spin and charge fluctuations as described below.

For nuclei with quadrupole moments, both a magnetic channel (electronic spin fluctuations) and a charge channel (local EFG fluctuations) contributes to the spin-lattice relaxation. $1/T_1 \propto \gamma^2$ for purely magnetic fluctuations, and $\propto Q^2$ for purely charge fluctuations [24,25]. From Tab. 1, $(^{79}\gamma/^{81}\gamma)^2 = 0.861$ and $(^{79}Q/^{81}Q)^2 = 1.389$. In the inset of Fig. 7, two horizontal lines, with values of $(^{79}Q/^{81}Q)^2$ and $(^{79}\gamma/^{81}\gamma)^2$, are plotted as references. At temperatures from 200 K down to 60 K, our measured $^{79}T_1^{-1}/^{81}T_1^{-1}$ data are very close to $(^{79}\gamma/^{81}\gamma)^2$, but far from $(^{79}Q/^{81}Q)^2$. This is consistent with dominant magnetic fluctuations when the systems is close to the magnetic ordering ($T_N \approx 74$ K). Below 60 K, the value of $^{79}T_1^{-1}/^{81}T_1^{-1}$ increases and moves toward $(^{79}Q/^{81}Q)^2$, which suggests enhanced contribution from charge fluctuations.

In the following, we decompose the $1/T_1$ into both a charge and a magnetic contribution,

$$T_1^{-1} = T_{1,s}^{-1} + T_{1,c}^{-1} \tag{2}$$

Taking

$$^{79}T_{1,s}^{-1} / \,^{81}T_{1,s}^{-1} = (^{79}\gamma / \,^{81}\gamma)^2 = 0.861 \tag{3}$$

and

$$^{79}T_{1,c}^{-1} / \,^{81}T_{1,c}^{-1} = (^{79}Q / \,^{81}Q)^2 = 1.389 \tag{4}$$

we obtain

$$^{79}T_{1,s}^{-1} = \frac{^{79}T_1^{-1} - 1.389 \times \,^{81}T_1^{-1}}{1 - 1.389/0.861} \tag{5}$$

$$^{79}T_{1,c}^{-1} = \frac{^{79}T_1^{-1} - 0.861 \times \,^{81}T_1^{-1}}{1 - 0.861/1.389} \tag{6}$$

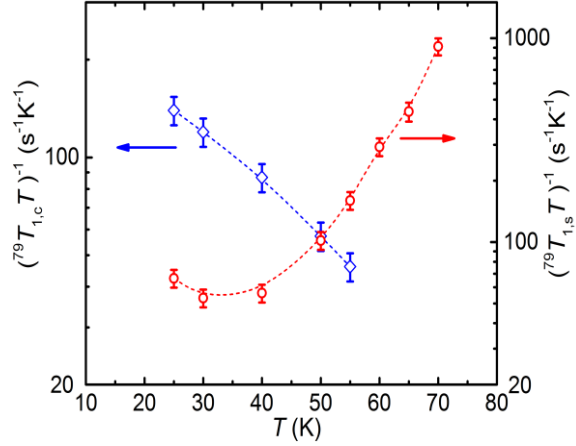

**Fig. 8.** The charge (left axis) and the spin (right axis) contribution to the $1/^{79}T_1T$, deduced from Fig. 7 (see text).

With equations 5 and 6, the $1/T_1$ data shown in Fig. 7 are decomposed into charge and spin contributions. In Fig. 8, the $1/^{79}T_1$, and $1/^{79}T_1$, are shown as functions of temperature. The charge contributions is nearly absent at temperatures above 60 K, and then arises rapidly upon further cooling. By contrast, the $1/^{79}T_1$, keeps decreasing with temperature below $T_N$. In particular, the $1/^{79}T_1$, exceeds $1/^{79}T_1$, below 40 K, indicating dominant charge fluctuations, rather than spin fluctuations.

## 7. Discussions

For all above measurements on high-quality $CuBr_2$ single crystals, our studies have revealed four prominent observations when cooled below ~60 K, which include (i) an enhancement of the $1/T_1$ in the charge channel, (ii) an increase of the FWHM of the NMR lines, (iii) a progressive decrease of $\varepsilon$, and (iv) a small increase of the magnetic susceptibility. Since all these behaviors onset at similar temperatures, the same physical origin is suggested. NMR as a low-energy probe, the increase of $1/T_1$, upon cooling is a clear evidence for enhanced low-energy EFG, or charge fluctuations.

These behaviors make large difference from charge and magnetic properties at high temperatures. For example, the low-energy spin fluctuations are dominated by magnetic fluctuations even with temperature up to 200 K. The $1/T_{1,c}$, on the other hand, becomes large far below the ferroelectric transition temperature, which suggests that the charge fluctuations are unlikely related to the spiral magnetic ordering and the ferroelectricity.

In fact, the increase of the magnetic susceptibility also indicates a suppression of spiral antiferromagnetic ordering with the onset of charge fluctuations. The decrease of the $\varepsilon$ below 60 K suggests that charge fluctuations compete with ferroelectricity. Furthermore, a small reduction of the zero-field NMR resonance frequencies were also observable below 10 K, as shown by $^{79}f$ and $^{81}f$ in the inset of Fig. 5, which suggest a reduction of the ordered magnetic moments. By contrast, the continuous increase of $1/T_1$, down to 10 K also supports a competition relation between charge fluctuations and the spiral magnetic ordering.

In the following, we describe possible origin for the charge fluctuations. In a crystal, lattice and electronic charge are combined to give the local EFG environment of nucleus. For quasi-1D and quasi-2D systems, charge fluctuations were reported with valence transition, orbital ordering, charge ordering, or CDW (charge-density-wave) transitions. In the current compound, since $Cu^{2+}$ has a stable $3d^9$ electronic configuration, any orbital ordering or valence fluctuations is unlikely to happen.

For quasi-1D magnetic compound, instabilities other than magnetic ordering, such as the spin-Peierls state and the charge ordering, can also arise. In particular, the spin-Peierls state is a popular ground state for antiferromagnet, where local spin singlets are formed among nearest neighbors. In this case, the unite cell is doubled with reduced atomic distance on the singlet bond, due to magnetoelectric coupling. When the system moves toward the spin-Peierls state, structural fluctuations can induce the large charge fluctuations. However, since the nearest exchange coupling $J_1$ is ferromagnetic in $CuBr_2$, the local singlet is unlikely to be formed among nearest neighbors. On the other hand, an incommensurate spin-Peierls state may be formed from the $J_1$-$J_2$ interactions. Recently, incommensurate spin-Peierls has been suggested in some quasi-1D material, such as $CuGeO_3$ [26] and TiOCl [27].

Charge ordering or charge density wave (CDW) states are also frequently reported in quasi-1D and quasi-2D materials, arising from strong Fermi surface nesting in low-dimensional systems. For example, charge ordering was reported in quasi-1D cuprates [28] or organic conductors $(TMTTF)_2X$ [29]. Recently, charge ordering has gained renewed interests in the cuprate superconductors [30] and the iron-based superconductors [31].

Both the spin-Peierls and the CDW state are in strong competition with the magnetic ordering. Our observation of reduced low-temperature dielectric constant seems to be consistent with such a competing picture. Furthermore, the large broadening of the NMR spectra is also consistent with the formation of an incommensurate state. Unfortunately, our NMR measurement is unable to differentiate these two candidate states, and other measurement techniques are strongly demanded. Recently, phonon modes below $T_N$ has been reported by a Raman study [22], although it was unclear whether it is caused by the magnetoelectric coupling or a competing mechanism with the spiral ordering. Our data suggests a competing charge fluctuations with the spiral ordering. Further investigation between the relations of these active phone modes and our observed charge fluctuations are strongly urged.

## 8. Summary

To summarize, we reported combined dielectric, susceptibility, NQR and zero-field NMR measurements on single crystals of a multiferroic compound $CuBr_2$. Below 60 K, we have resolved the zero-field NMR $^{79}$Br and $^{81}$Br spectra, and found enhanced charge fluctuations from the spin-lattice relaxation rates. The charge fluctuations are accompanied with the NMR linewidth broadening, the increase in the magnetic susceptibility, and the reduced dielectric constant. Since the high quality of the samples has been demonstrated by the sharp magnetic transition at $T_N \approx 74$ K, our data suggest emergent charge fluctuations with an intrinsic origin, which affects both the charge and the magnetic properties of the system, but in competition with the spiral magnetic ordering and the ferroelectricity. Other measurement techniques are requested to reveal the nature of such a competing instability.